\documentclass[epjCONF]{svjour}
\usepackage{graphics}
\usepackage[varg]{txfonts} % Times fonts
\usepackage[latin1]{inputenc}
\newcommand{\teff}{\mbox{${T}_{\rm eff}$}}

\newcommand{\rsun}{\mbox{${\rm R}_{\odot}$}}
\newcommand{\msun}{\mbox{${\rm M}_{\odot}$}}
\newcommand{\lsun}{\mbox{${\rm L}_{\odot}$}}

\newcommand{\LS}{\mbox{$\langle \Delta \nu\rangle$}}

\newcommand{\DPa}{\mbox{$\langle \Delta P \rangle_a$}}
\newcommand{\BV}{Brunt-V\"ais\"al\"a}
\newcommand{\numax}{\mbox{$\nu_{\rm max}$}}

\newcommand{\simgt}{\lower.5ex\hbox{$\; \buildrel > \over \sim \;$}}
\newcommand{\simlt}{\lower.5ex\hbox{$\; \buildrel < \over \sim \;$}}

\def\apj{ApJ~}%
\def\apjl{ApJ~}%
\def\apjs{ApJS~}%
\def\aap{A\&A~}%
\def\mnras{MNRAS~}%
\def\solphys{Sol.~Phys.~}%
\def\nat{Nature~}%
\def\memsai{Mem.~Soc.~Astron.~Italiana~}%
\session-title{Ageing low mass stars: from red giants to white dwarfs}
\begin{document}
\title{Asteroseismology of red giant stars: the potential of dipole modes}
\author{J. Montalb{\'a}n\inst{1}\fnmsep\thanks{\email{j.montalban@ulg.ac.be}}
  \and A. Noels\inst{1} }

\institute{Universit{\'e} de Li{\`e}ge, All{\'e}e du Six Ao\^ut -17, B-4000  Li{\`e}ge, Belgium}

\abstract{Since the detection of non-radial solar-like oscillation modes in red giants with the CoRoT satellite, the interest in the asteroseismic properties of red giants and the link with their global properties and internal structure is increasing. Moreover, more and more precise data are being collected with
 the space-based telescopes CoRoT and {\it Kepler}. Particularly relevant has been  the detection of  mixed modes in a large number of G-K red giants. In this contribution we discuss the potential of these dipole mixed modes to provide information on core extra-mixing and transport of angular momentum.} 
%end of abstract
%
\maketitle

\section{Introduction}
\label{sec:intro}
Red giants are cool stars with an extended and diluted convective envelope that, as in solar-like stars, can stochastically excite oscillation modes. Their frequency spectra are characterized by two global parameters:  the frequency at maximum power   (\numax) and the frequency separation between consecutive radial modes (\LS), which are linked, via scaling relations, to global stellar parameters: mass, radius and effective temperature (\teff) \cite{kb1995}. The detailed properties of the oscillation modes, however, depend on the stellar structure. Because of the contraction of the inert He core (which increases the frequency of gravity modes) and of the expansion of the hydrogen rich envelope (which involves a drop of mean density and hence of the frequency of pressure modes), modes with frequencies in the solar-like domain can propagate in the gravity and in the acoustic cavities, presenting hence  a mixed gravity-pressure character. So, while the solar-like spectra of main sequence (MS) pulsators are mainly made up of a moderate number of acoustic modes by angular degree, those of red giants can present, in addition to radial modes, a large number of non-radial g-p mixed modes.

The dominant p- or g-character of these non-radial modes depends on the cavity where they mainly propagate,  and may be estimated from the value of the normalized mode inertia ($E$, see e.g. \cite{jcd2004} and references therein). Modes trapped in high-density regions (g modes) have high  $E$, while pure p modes such as the radial ones have the lowest $E$. Depending  on the coupling between the two cavities, some non-radial modes  may be well trapped in the acoustic cavity  and behave as p modes presenting an inertia close to that of the radial modes, while modes with strong mixed g-p character have larger $E$. Several observational and theoretical studies based on pressure-dominated modes in red giants have been published in the last few years: behavior of frequency patterns such as large and small frequency separations, and their relation with the corresponding asymptotic quantities \cite{carrieretal2010,beddingetal2010,huberetal2010,montalbanetal2010a,montalbanetal2010b,montalbanetal2012}; detection of sound speed variations and the relation with the second ionization zone of He \cite{miglioetal2010}; properties of the echelle diagrams in the form of folded echelle diagrams and "universal patterns" \cite{beddingkjeldsen2010,mosseretalUniversal2011}.

Assuming the inertia of the mode as a good proxy of its amplitude, mixed modes with not too large value of $E$ should be observable. In fact, as observation time of red giants increased, the frequency resolution noticeably improved and it became possible not only to detect the acoustic modes in the spectra of red giants, but also to distinguish a forest of mixed modes around each of them \cite{becketal2011}. The frequency (or period) separation between these modes depends on the evolutionary state  and on the mass of the model. These properties have been measured by \cite{beddingetal2011} in the spectra of red giants observed by {\it Kepler} and also in those of CoRoT red giants \cite{mosseretal2011}, and the comparison with theoretical computations allowed them to establish  a relation between the value of this period separation and the evolutionary state of the star, lifting the degeneracy between red giant branch (RGB) and central He-burning (He-B). Finally, an additional structure has been observed in the spectra of dipole modes of some red giants. It has been identify as rotational splitting and used to derive for the first time the internal rotation profile in stars different from the Sun \cite{becketal2012}. 

For a given red giant model, the dipole modes have a stronger mixed character than the quadrupole ones. Since the coupling between the acoustic and gravity cavities is more important for $\ell=1$ modes, they are the best tools to access to the inner regions of the star.  In this contribution  we focus mainly on the potential of dipolar modes to provide information on the evolutionary state, and also on the transport processes undergone during the main sequence and the He-burning phases. 
In section~\ref{sec:evolution} we remind some relevant aspects of the internal structure and evolution of red giants for the  understanding of  their seismic properties.  The latter are described in section~\ref{sec:deltap}. The potential of dipole mode to provide information on core extra-mixing and on internal rotation is discussed in sections~\ref{sec:over} and ~\ref{sec:rotation} respectively.

\section{Red giants: internal structure and evolution}

\label{sec:evolution}

\begin{figure}
\resizebox{\hsize}{!}{{\includegraphics{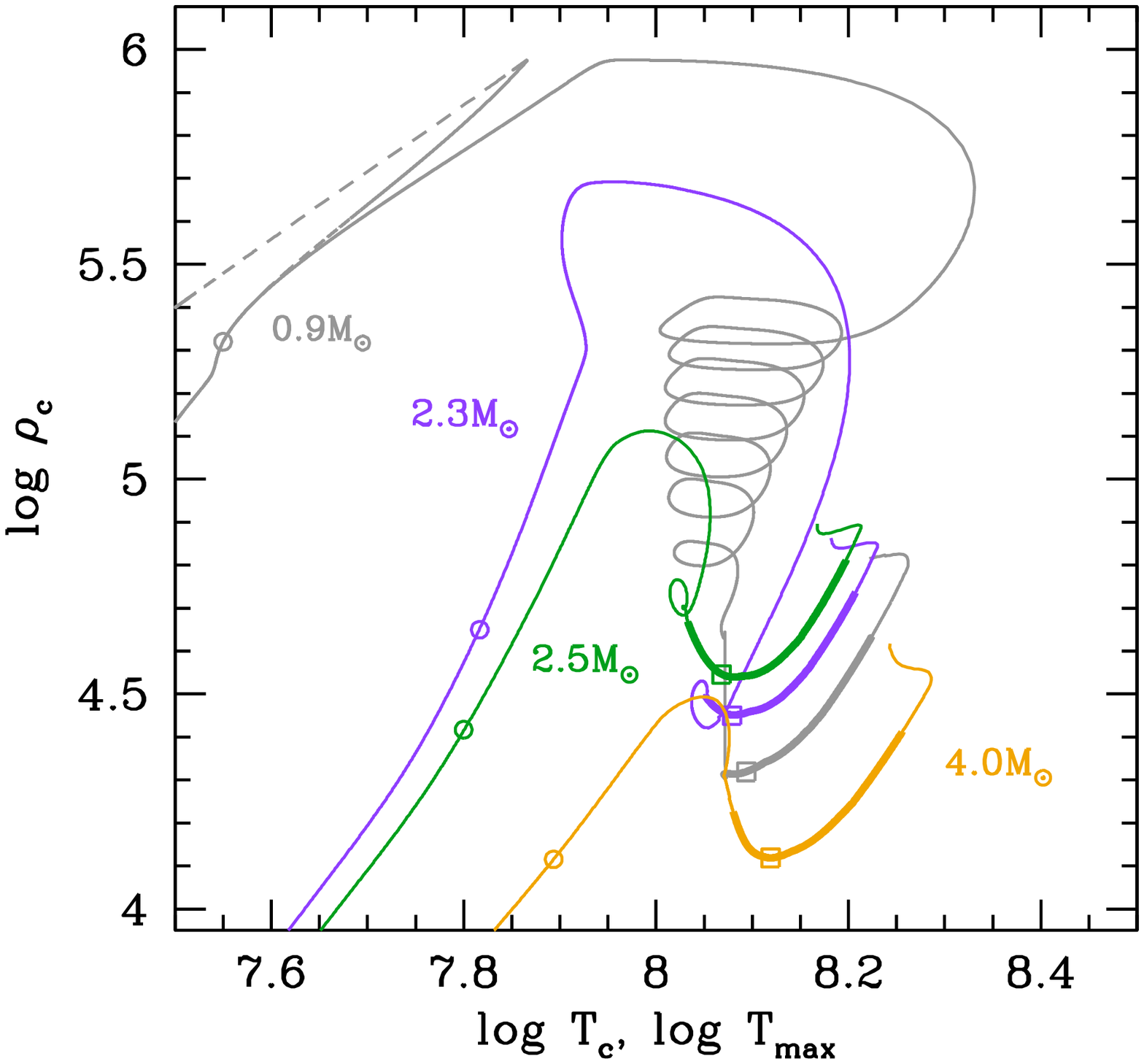}}{\includegraphics{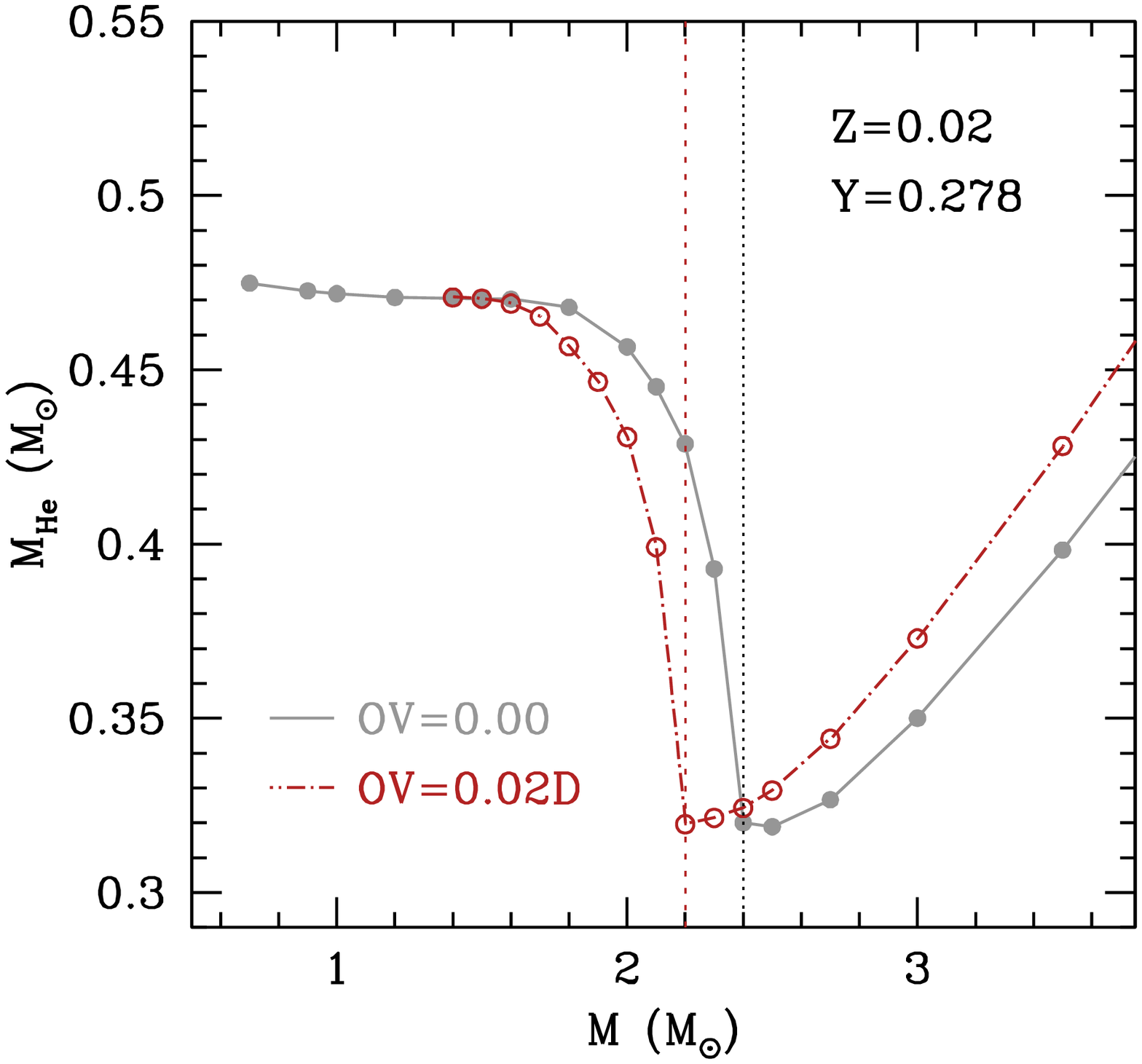}}}
\caption{Left panel: Evolution of $\rho_{\rm c}$ and $T_{\rm c}$ (dashed line) and $T_{\rm max}$ (solid line)  from the RGB phase up to the exhaustion of helium at the center for masses lower and larger than $M_{\rm tr}$ (2.4~\msun\ for the chosen chemical composition and core extra-mixing). Squares corresponds to models with $Y_{\rm c}=0.5$, and the open circles represent the RGB models with the same radius as the He-B-$Y_{\rm c}=0.5$ ones. Right panel: He-core mass at the He ignition as a function of the total mass for models without (solid lines) and with (dash-dotted line) overshooting, and chemical composition $Z=0.02$, $Y=0.278$. $M_{\rm tr}$ changes from 2.4~\msun\ to  2.2~\msun\$ (vertical lines).} 
\label{fig:flash} 
\end{figure}

A detailed  description of the red giant evolution phase can be found for instance in \cite{salariscassisi2005} and in the Weiss, Salaris and Bressan contributions in this volume. In this section we focus on the post-MS evolution of low and intermediate mass stars seen from their center thermodynamical conditions ($\rho_{\rm c}$ and $T_{\rm c}$), from the climb up the RGB till the exhaustion of central He ($Y_{\rm c}$) as shown Fig~\ref{fig:flash} (left panel). The central He-burning phase corresponding to $0.9\geq Y_{\rm c} \geq 0.05$ is highlighted with a thicker line.   For all these stars the inner contraction during the RGB phase leads to an increase of the density and the temperature, until the latter reaches the value required to ignite He  by the 3$\alpha$ nuclear reaction ($\sim10^8$~K), ending then the RGB phase; a readjustment process of the thermodynamical properties follows the onset of He-burning until nuclear reactions reach the equilibrium. The details of the path followed depends on the stellar mass, and the drop in central density can vary from a factor 50 for low mass stars (i.e. 0.9~\msun) to a factor 2.5 for 4~\msun.

In stars with $M \geq 2.5$~\msun, $3\alpha$ nuclear reactions ignite at the center of a non-degenerate core, hence the pressure regulation mechanism can immediately take place. The core expands, which prevents any heating whose result would be an uncontrolled increase of the nuclear energy production rate. 
 
Stars with $M\leq 1.8$ follows very similar paths in the plane ($\rho_{\rm c},T_{\rm c}$) and the curve of the 0.9~\msun\ model is representative of the behavior of low mass models. In these low-mass stars, the density becomes extremely large during the RGB ascension and the level of degeneracy increases accordingly. The release of gravitational energy from the contracting helium core can no longer increase the temperature. Moreover, electron conduction becomes the major source of opacity while an additional cooling is provided by neutrinos. On the contrary the region below the non degenerate H-shell heats up and an inverted temperature profile is formed, with a maximum ($T_{\rm max}$) reached at $M_r\sim 0.2$~\msun. The dashed line in Fig.~\ref{fig:flash} represents the evolution of $T_{\rm c}$, while the solid line correspond to that of $T_{\rm max}$. When the He-core mass is about 0.48~\msun, $T_{\rm max}$  reaches $\sim10^8$~K and He ignites off-center. Since in degenerate conditions the pressure is not sensitive to temperature changes, the initial temperature rise induced by nuclear reactions is not immediately reversed by the expansion and cooling of the affected layers. The temperature continues rising and so does the nuclear reaction rate. A huge amount of energy ($L\sim 10^{10}$~\lsun) is produced in a very short time interval. Its  absorption by the surrounding layers lifts  the degeneracy in the layers between $M_r\sim 0.2$ and the outer edge of the He-core. As a result, the pressure regulation mechanism can act, decreasing the density and the  temperature and turning off the nuclear reactions. A new cycle of contraction/heating begins, and the degeneracy is progressively lifted in layers closer and closer to the center through a sequence of secondary flashes (loops in the 0.9~\msun\ curve of Fig.~\ref{fig:flash}) until He-burning eventually reaches the center. 
 
As the stellar mass increases, the maximum central density and level of degeneracy reached during the RGB ascension decrease and so does the He-core mass required to ignites He. For 2.3~\msun\, the central temperature still decreases close to the tip, but $T_{\rm c}=T_{\rm max}$. The He-flash thus occurs at the center and its strength is two orders of magnitude smaller than for 2.1~\msun.  This transition between off-center and center He-flash is very abrupt (0.1~\msun) and occurs close to  the transition mass ($M_{\rm tr}$) between low and intermediate mass stars.

As shown in Fig.~\ref{fig:flash} (right panel), stars with $M \lessapprox 2.0$~\msun\ have the same He-core mass $\sim 0.48$~\msun\ at the He ignition. This value rapidly decreases as the stellar mass increases and, after reaching a minimum value ($\sim 0.33$~\msun, the lowest mass of a pure helium star for stable He-burning--see e.g. \cite{kippenhahn1990}) at the transition mass, it increases again following the total mass (i.e., $M_{\rm He}\sim 0.1$~M$_T$), as a result of the larger and larger convective core mass during the MS phase (\cite{girardietal1998,castellanietal2000} and references therein). He-burning stars with masses smaller than $M_{\rm tr}$ populate the so-called  Red Clump (RC, \cite{girardietal1998}), while  stars with mass close to the transition one form the so-called ``secondary clump" \cite{girardi1999}. $M_{\rm tr}$ depends on the extent of MS  extra-mixing, and the same is true for the age and the mass of the stars populating the secondary clump.

 \begin{figure}
\resizebox{0.55\hsize}{!}{{\includegraphics{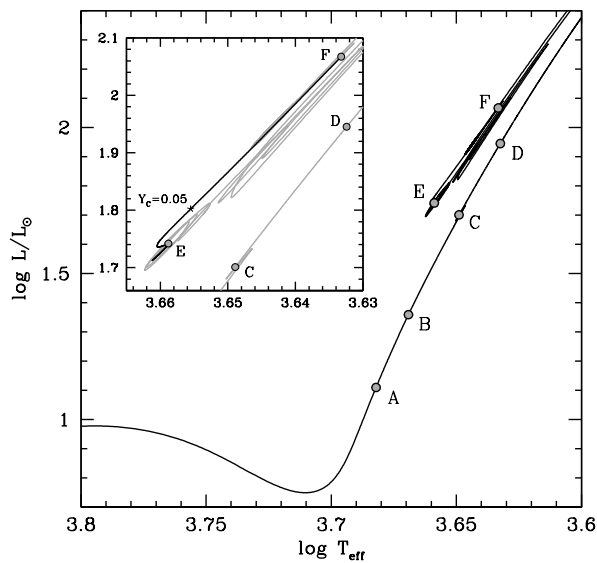}}}
\hspace*{0.5cm}
\begin{minipage}[t]{0.40\linewidth}
\vspace*{-7.5cm}
\caption{Evolutionary track of a 1.5~\msun star in the red giant phase, RGB and central He-burning. Dots and labels mark location of models discussed in the text and in Figs.~\ref{fig:spec},~\ref{fig:echelle} and \ref{fig:rot}. In particular ``E'' is the model with $Y_{\rm c}=0.8$ and ``F'' the model just before the He-shell development, with $Y_{\rm c}=5.\times 10^{-4}$, and the asterisk corresponds to the model with $Y_{\rm c}=0.05$. The values of stellar radius and density contrast ($R/R_{\odot}$, $\rho_{\rm c}/\langle\rho\rangle$) for the models A--F are: A: (5.2, $7.0\times10^6$), B: (7.3, $2.6\times10^7$), C: (11.9, $1.8\times10^8$), D: (17.0, $6.2\times10^8$), E: (11.9, $1.7\times10^7$) and F: (19.5, $2.0\times10^8$).} 
\label{fig:hr} 
\end{minipage}
\end{figure}

\begin{figure}
\resizebox{\hsize}{!}{{\includegraphics{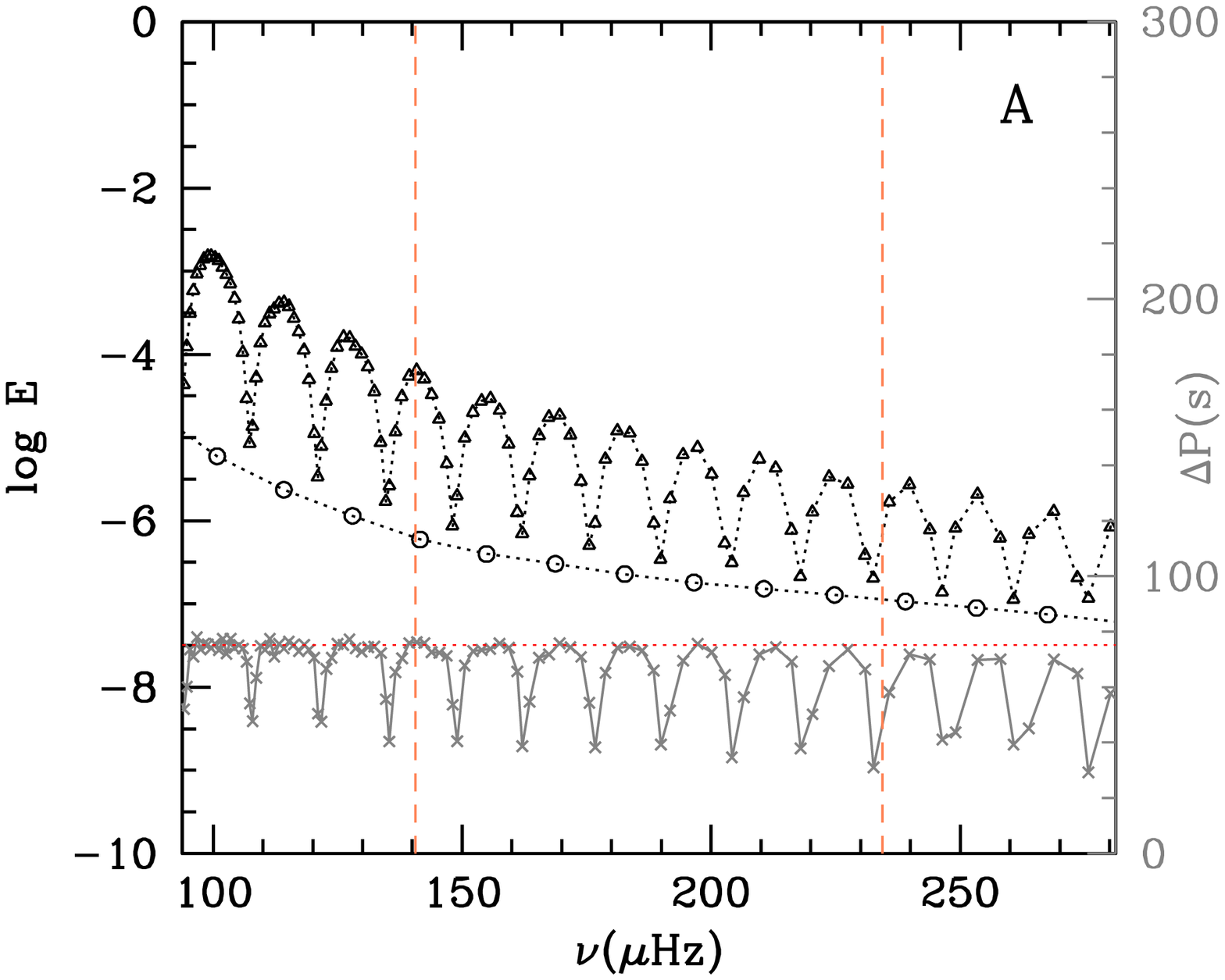}}{\includegraphics{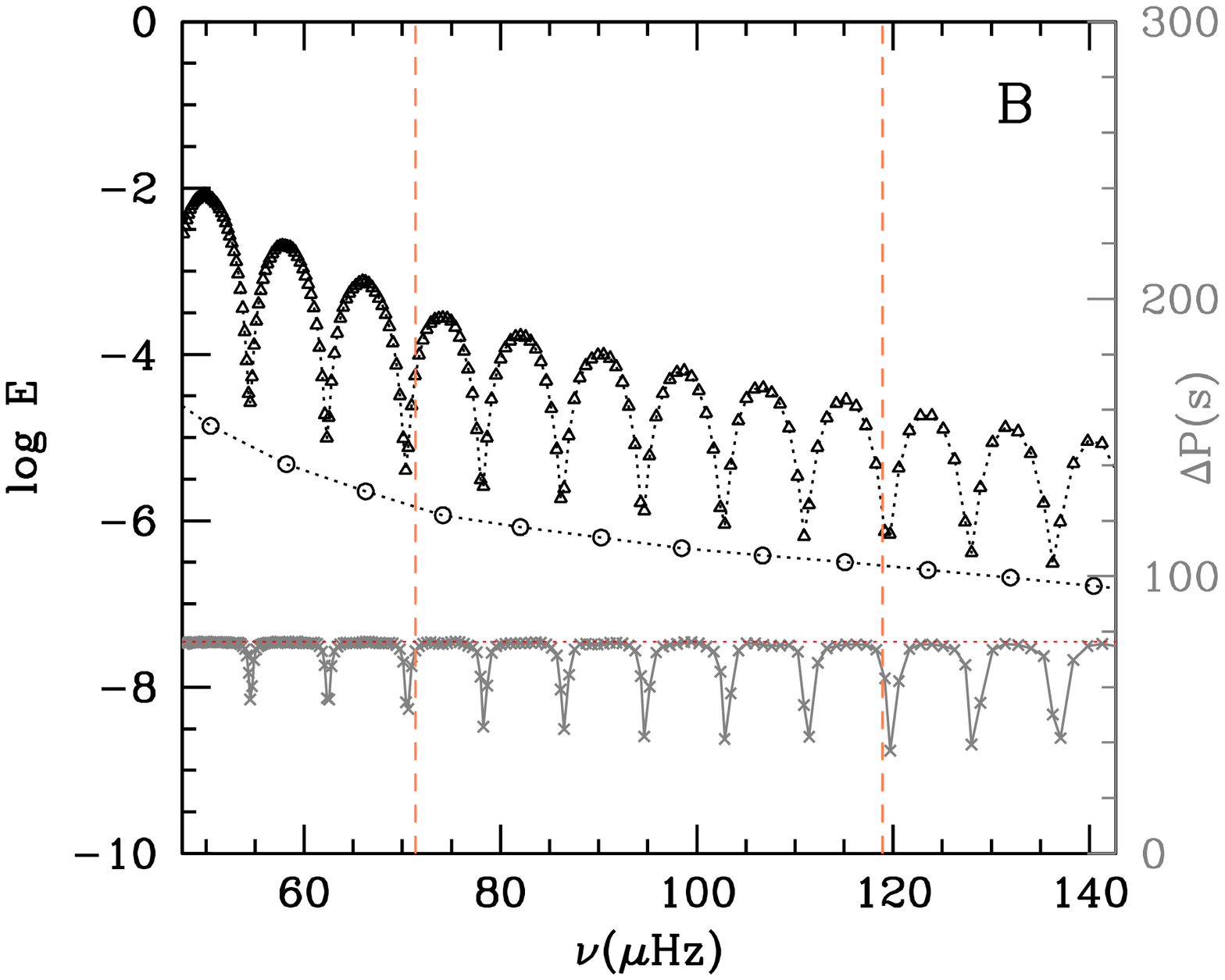}}}
\resizebox{\hsize}{!}{{\includegraphics{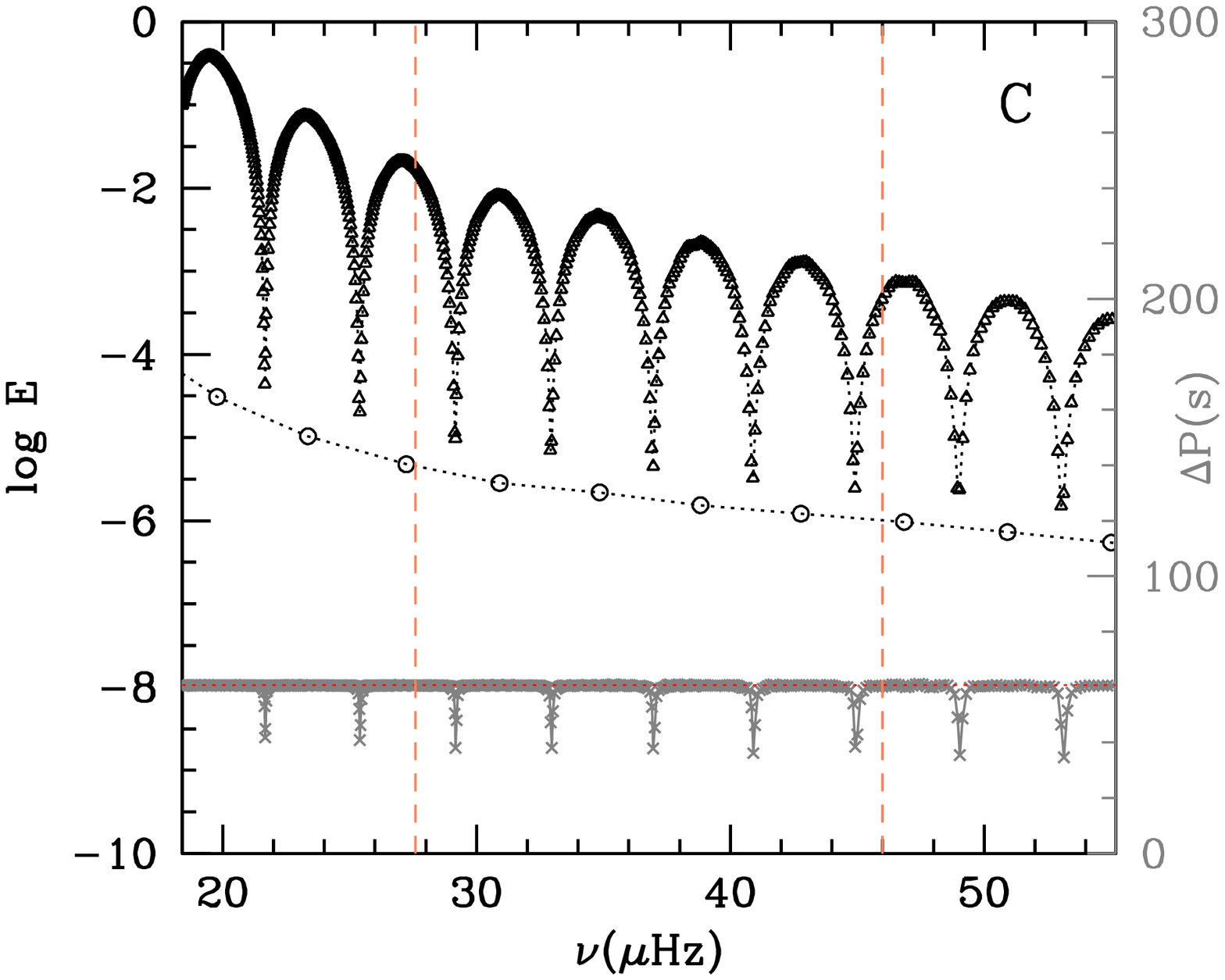}}{\includegraphics{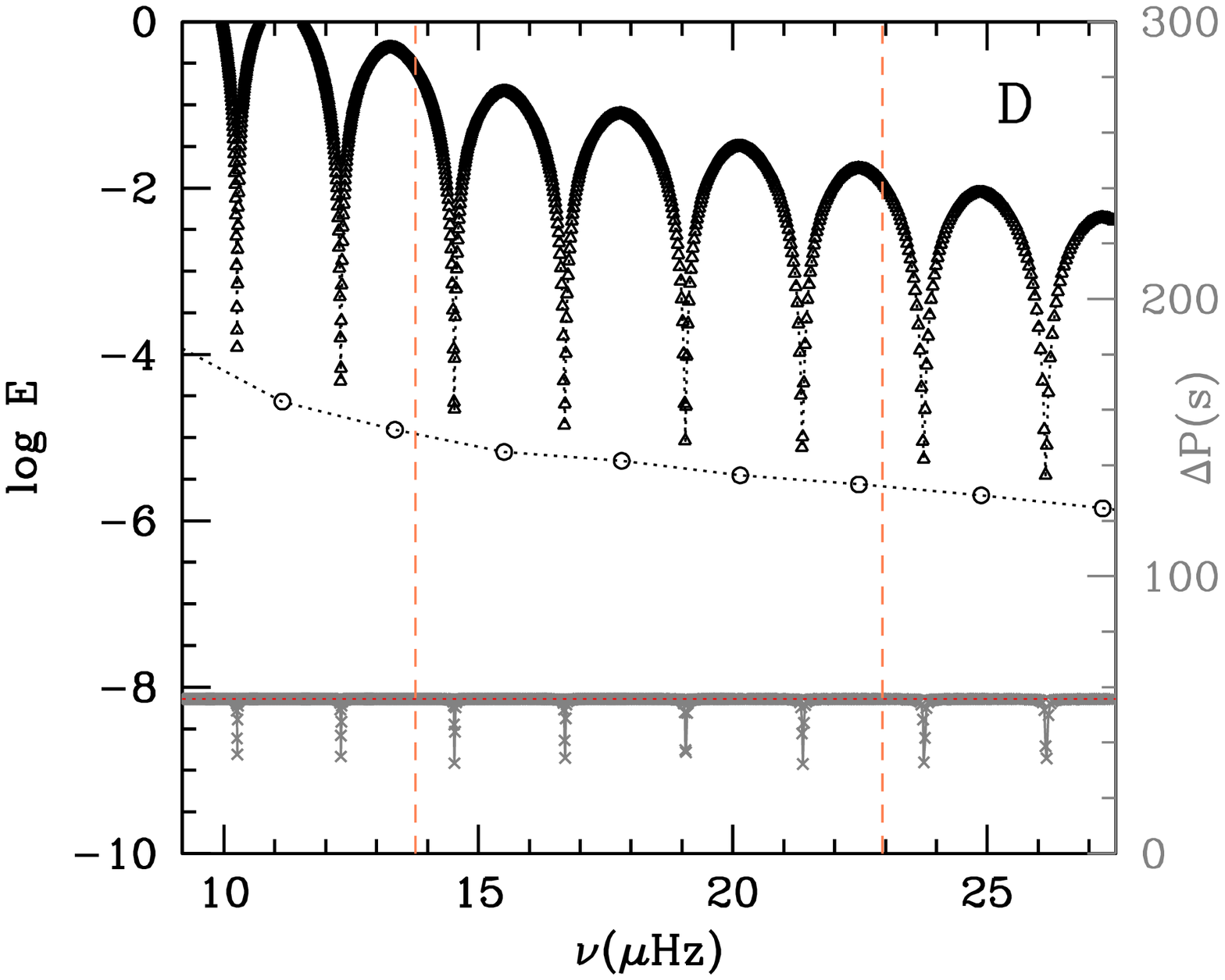}}}
\resizebox{\hsize}{!}{{\includegraphics{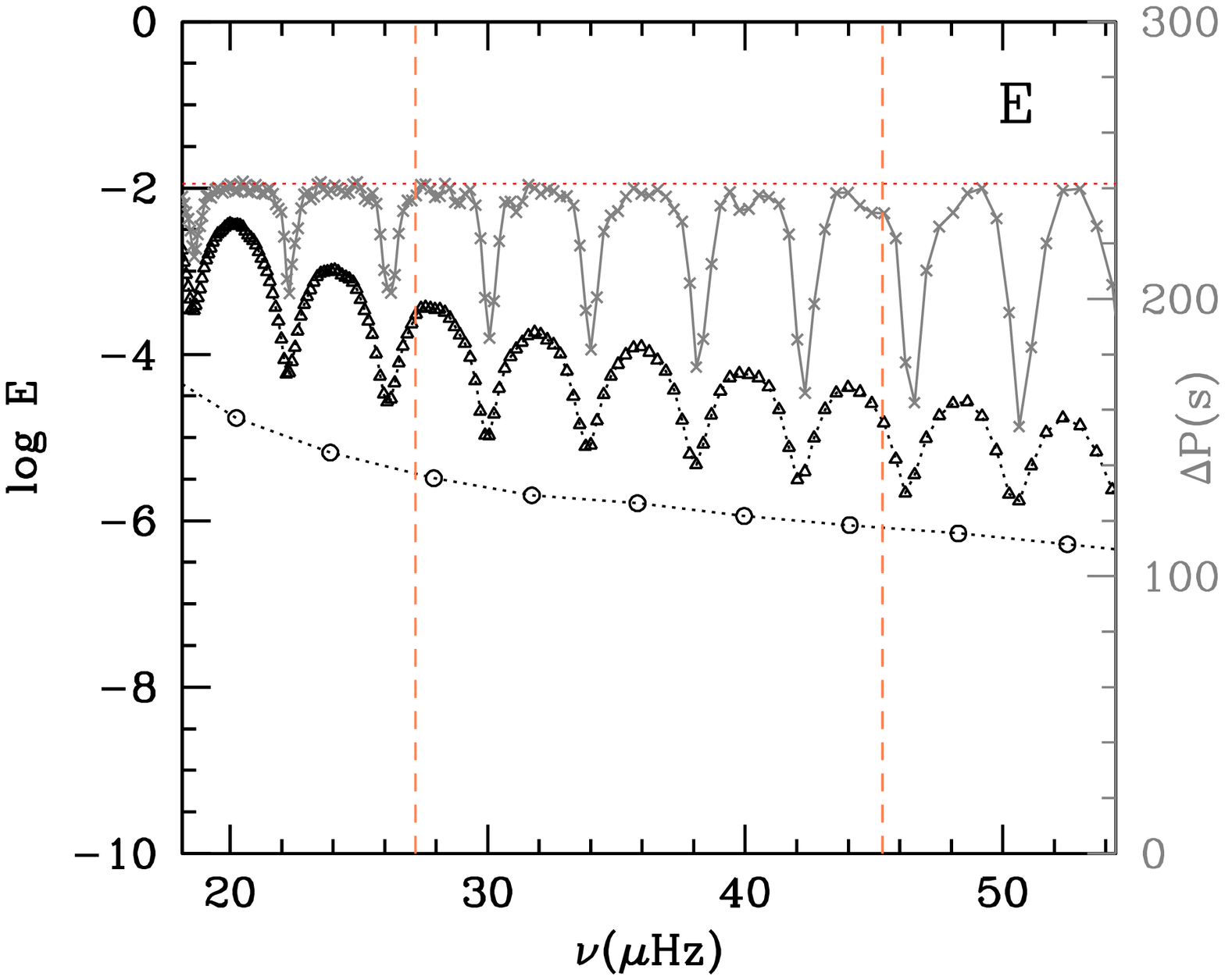}}{\includegraphics{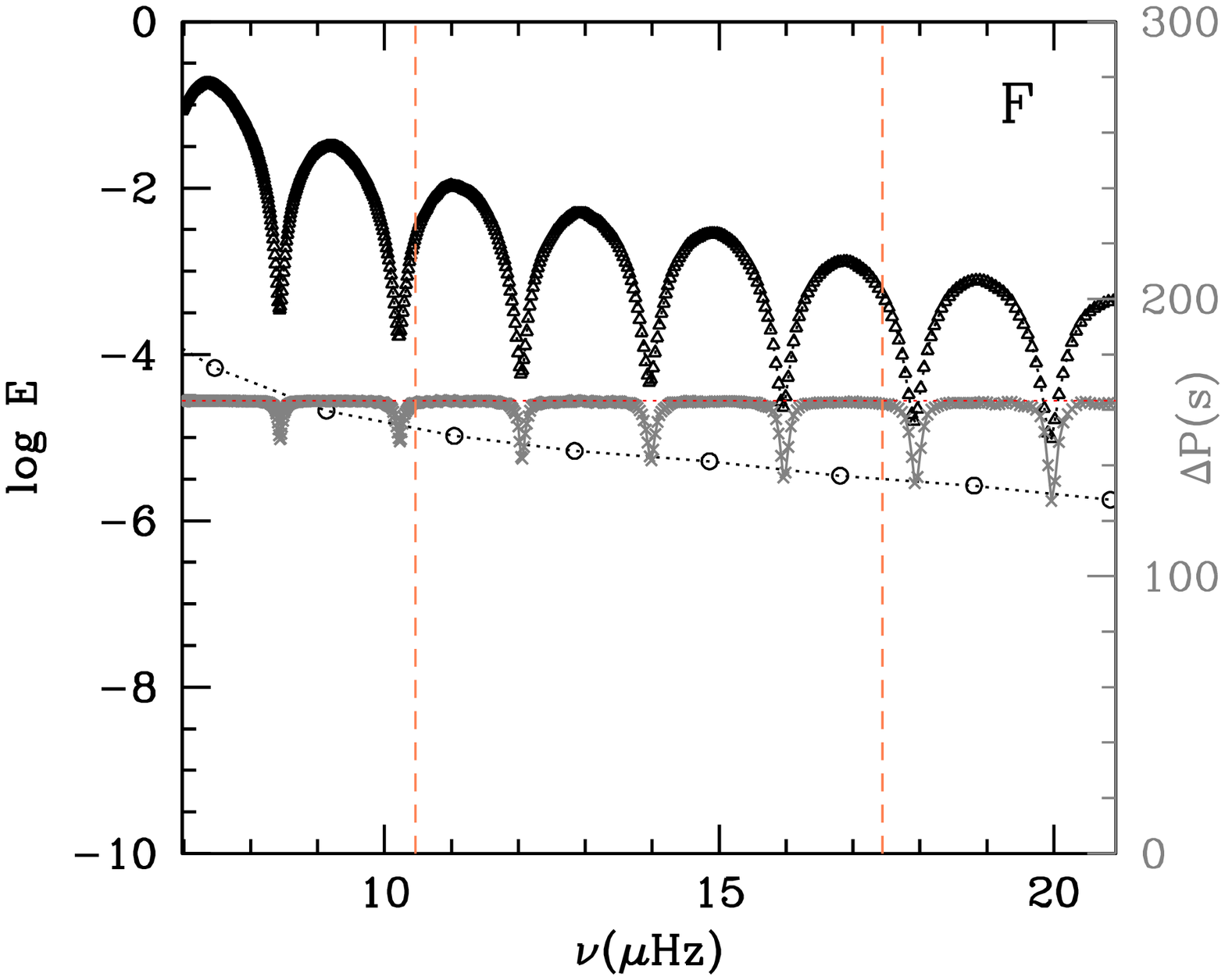}}}
\caption{ Panels A--F: Oscillations spectra for models A--F in Fig.~\ref{fig:hr}. In each panel we plot the mode inertia ($E$) {\it versus} frequency for radial ($\ell$=0, circles) and dipole ($\ell=1$,
triangles) modes. Vertical dashed lines denote the frequency domain of solar like oscillations. 
Grey crosses and lines represent the period separation (right axis) between consecutive dipole modes {\it versus} frequency, and horizontal thin-dashed lines indicate the value of the corresponding asymptotic period spacing $\langle\Delta P\rangle_{\rm a}$.}
\label{fig:spec}
\vspace*{-0.85cm}
\end{figure}

Following the readjustment, the evolution of $\rho_{\rm c}$ and $T_{\rm c}$ during the quiet He-burning phase (He-B MS, thicker line) for models with $M \leq 2.3$~\msun\ is quite similar. Stars with $M \leq 1.8$~\msun\ all reach the He-B phase at the same point in the plane ($\rho_{\rm c}$,$T_{\rm c}$) and follow exactly the same evolution curve regardless the total mass, indicating that the evolution of the central region depends only on the He-core mass and is almost decoupled from that of the envelope, which is driven by the H-burning shell. In the interval $1.8\leq M(M_\odot)\leq 2.3$ the tracks in the plane $\rho_{\rm c},T_{\rm c}$ follow quasi parallel trajectories during the He-B phase and occupy the space between the 0.9 and 2.3~\msun\ curves. For all of them, $\rho_{\rm c}$ and $T_{\rm c}$ increase as the central  mass fraction of He ($Y_{\rm c}$) decreases.

For stars that ignite He at the center ($M \geq 2.3$~\msun) the core expansion continues during a fraction of the central He-burning until the readjustment phase ends. Afterwards, $\rho_{\rm c}$ and $T_{\rm c}$ increase as $Y_{\rm c}$ decreases. When $Y_{\rm c} \leq 10^{-4}$, central He-burning stops, the degenerate CO core contracts and cools down while He-burning is shifted into a shell surrounding the CO core. The star enters the Early AGB phase.

\begin{figure}
\resizebox{\hsize}{!}{{\includegraphics{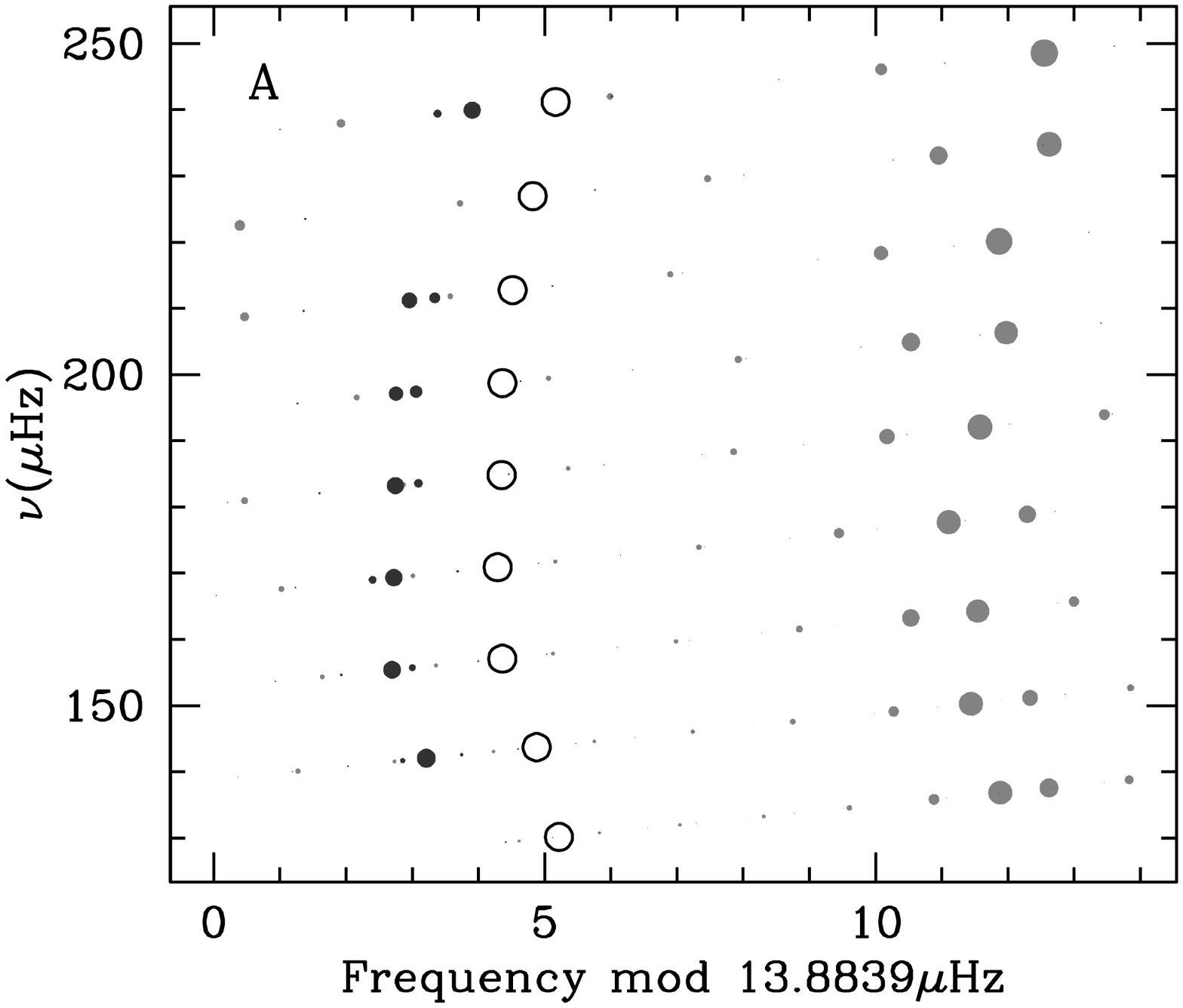}}{\includegraphics{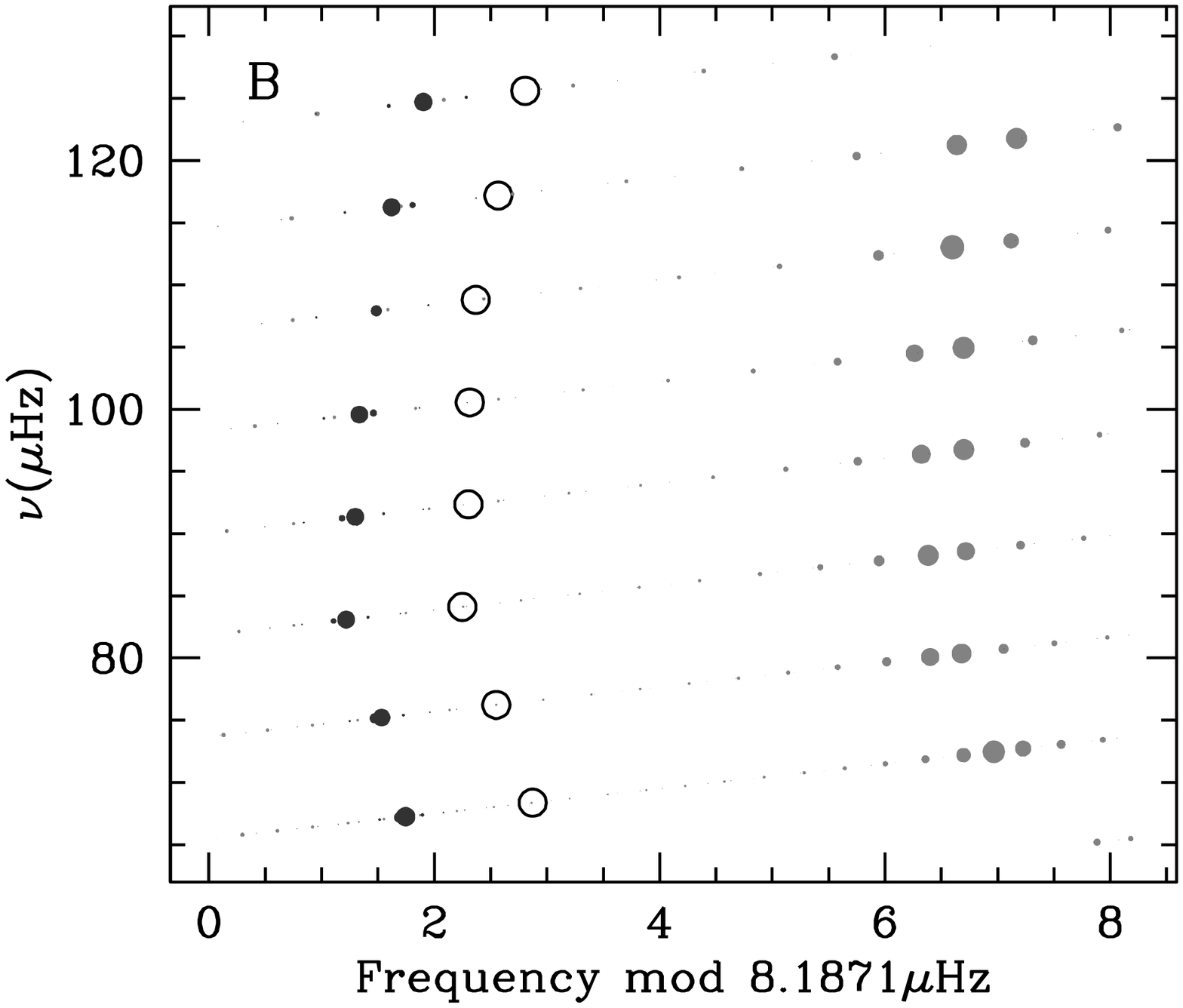}}}
\resizebox{\hsize}{!}{{\includegraphics{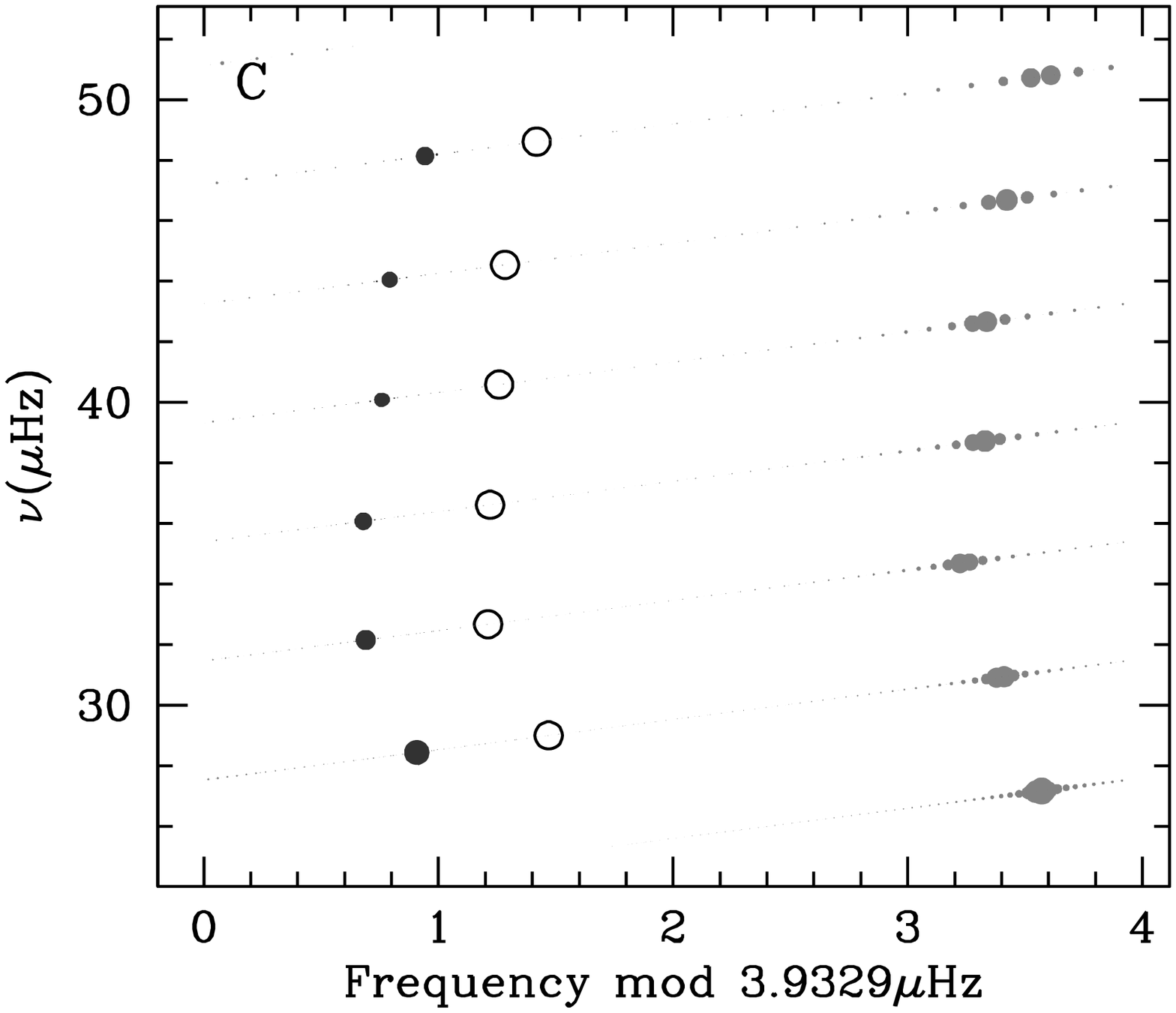}}{\includegraphics{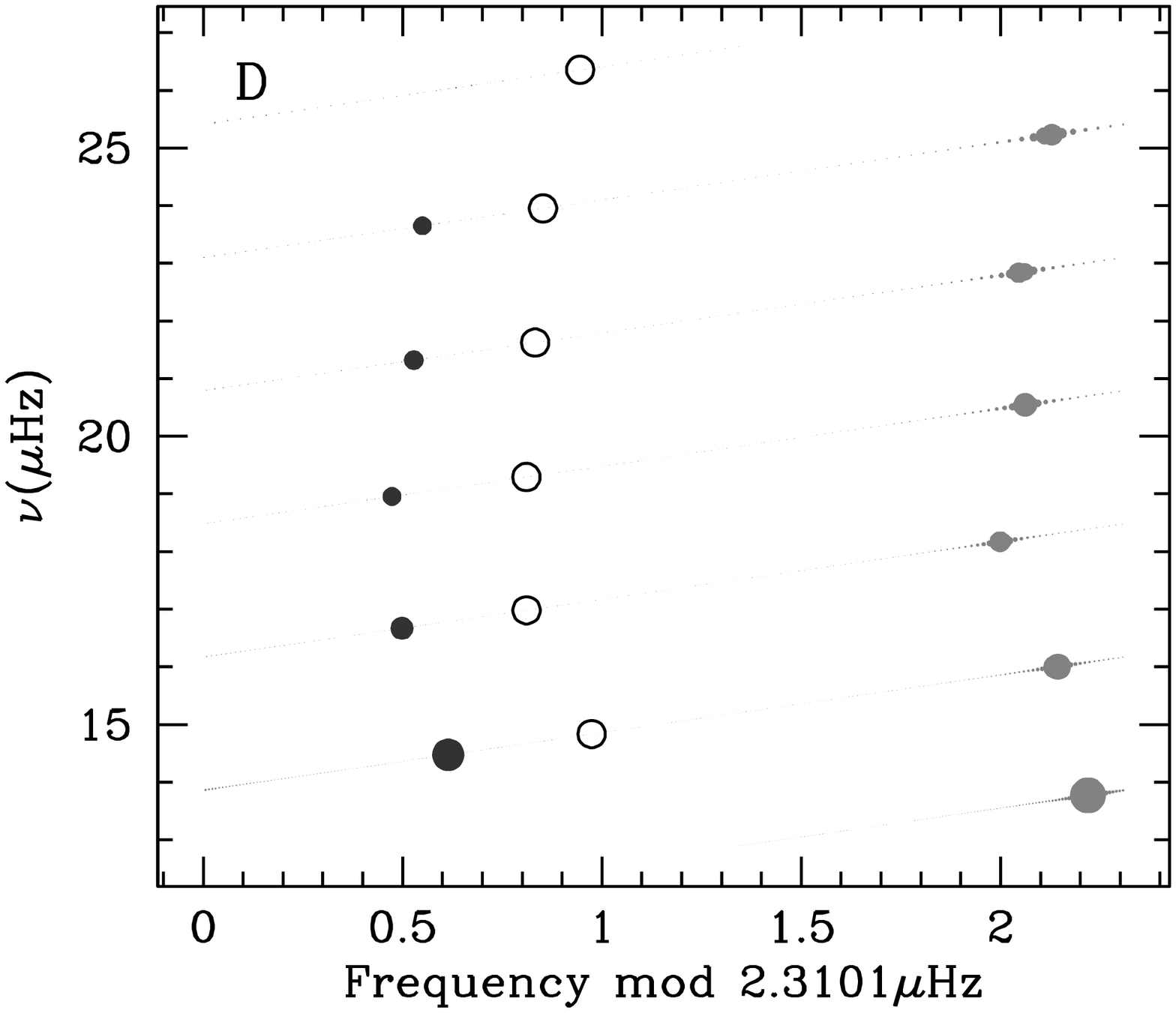}}}
\resizebox{\hsize}{!}{{\includegraphics{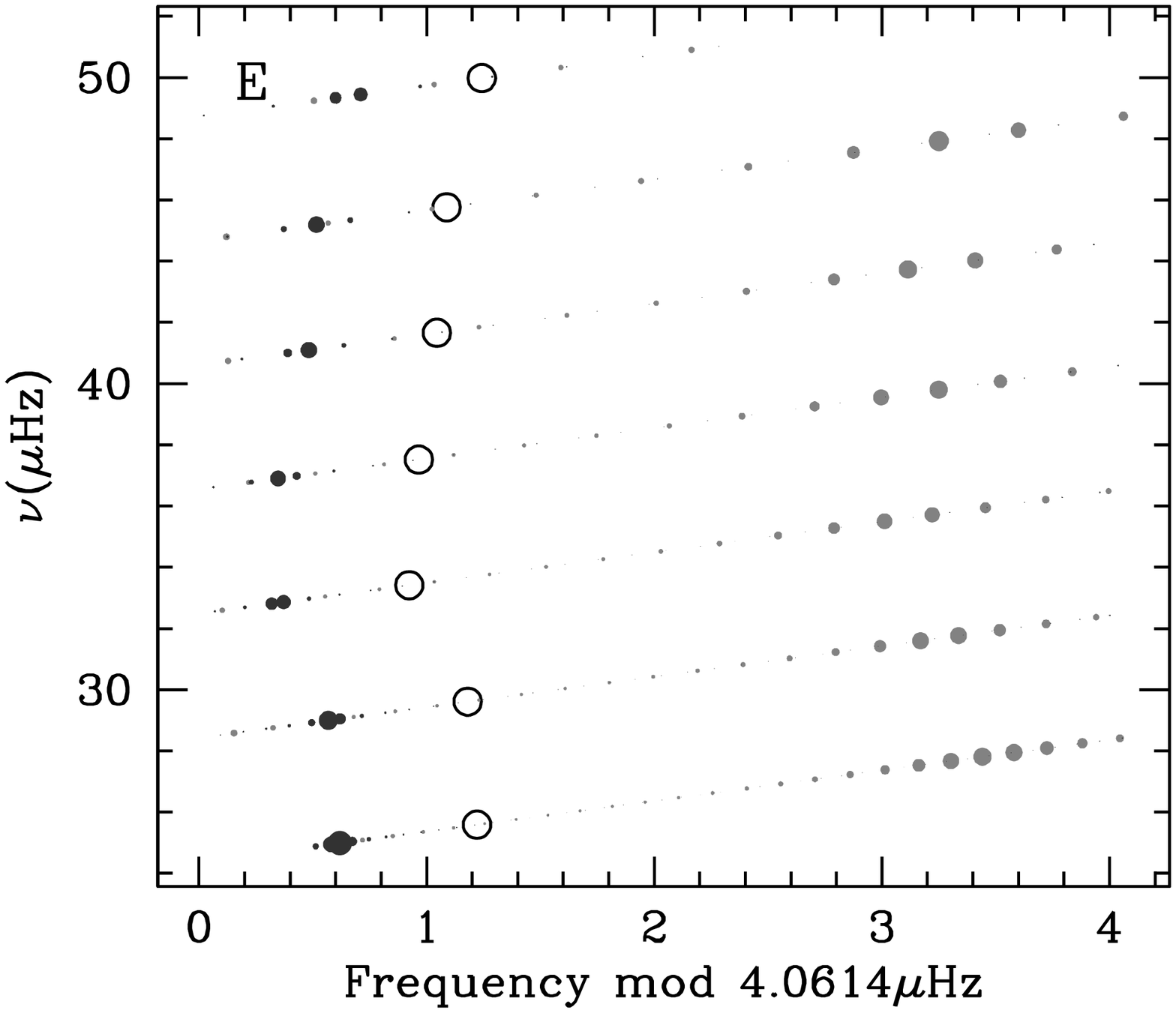}}{\includegraphics{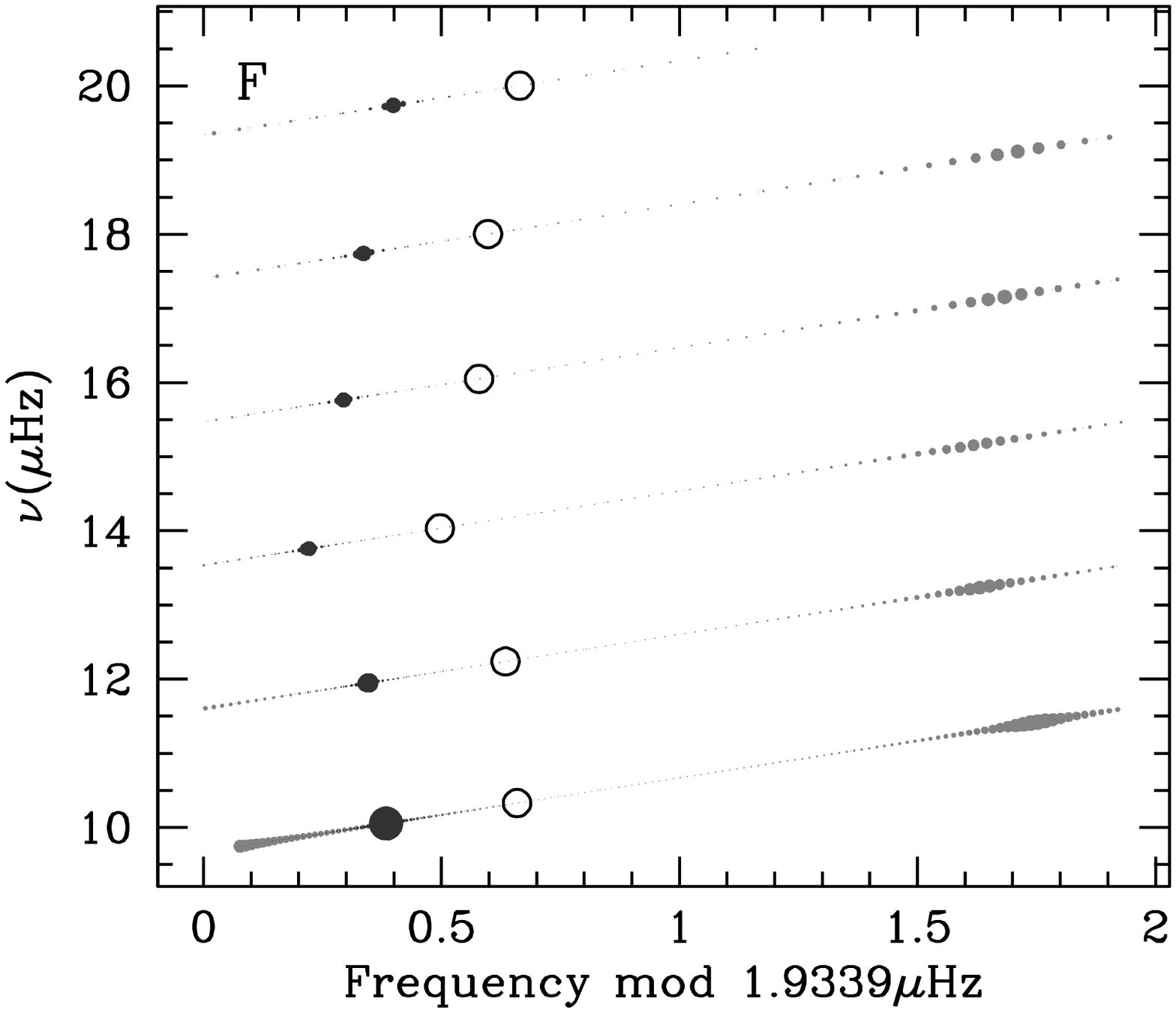}}}
\caption{Expected echelle diagram for the oscillation spectra of models A--F in Fig.~\ref{fig:hr}. Open circles correspond to $\ell=0$ modes, dark-solid dots to $\ell=2$ ones, and grey-solid dots to dipole modes. The symbol size is related to the relative amplitude taking as reference that of radial modes (see the text).}
\label{fig:echelle}
\end{figure}

\section {Adiabatic oscillation properties: period spacing}
\label{sec:deltap}

The properties of oscillation spectra depend on the profile of the \BV\ ($N$) and Lamb ($S_\ell$)  frequencies.  As mentioned above, because of the high density contrast between the core and the envelope,  the RG oscillation spectrum is characterized by  a large number of mixed p-g modes in addition to the radial ones. The asymptotic approximation \cite{tassoul1980} predicts that the number of  g-modes by radial order is: $n_g\propto (\ell\,(\ell+1))^{1/2}\int N/r dr$, and that the periods  of two modes of same degree ($\ell$) and consecutive order ($n$)  are separated by a constant value \DPa:

\begin{equation}
\langle \Delta P \rangle_{\rm a} = \frac{2 \pi^2}{\sqrt{\ell (\ell+1)}} \frac{1}{\int N/r dr}.
\label{eq_dp}
\end{equation}

As the star climbs the RGB (A--D in Fig.~\ref{fig:hr}) and the inert helium core contracts, the peak of $N$ (corresponding to the H-shell) shifts more and more inwards.  As a consequence $n_g$ increases (and the period spacing decreases) as the star goes from the bottom to the tip of the RGB. Once the temperature for He ignition is reached,  the onset of nuclear burning  is accompanied by the expansion of the star  central regions and the development of a convective region. In the He-B model the expansion of the central layers  leads to a lower $N$  maximum, and its location is displaced at a larger distance from the center. Moreover, the central convective regions do not contribute to the integral in Eq.~\ref{eq_dp}. Both effects act in the same sense, decreasing $n_g$ and increasing \DPa.  Models C and E (Fig.~\ref{fig:hr}) have the same mass and radius (and hence the same \numax\ and \LS), but the density contrast  ($\rho_{\rm c}/\langle \rho\rangle$) has decreased by a factor 10 between those evolutionary stages. Moreover model E has developed a small convective core. As a consequence, the period spacing between consecutive g-modes, as determined from Eq.~\ref{eq_dp}, is significantly smaller in the RGB model ($60$~s) than in the He-B one ($240$~s). 

Each panel of Fig.~\ref{fig:spec} shows the properties of dipole modes for the corresponding models in Fig.~\ref{fig:hr}. We plot the mode inertia of radial (open circles) and dipole (triangles) modes, as a function of frequency, as well as the period separation  between dipole modes of consecutive radial order ($\Delta P=P(n+1)-P(n)$). The value of \DPa\ is indicated by an horizontal dotted line. 

Between models A and D, the stellar radius changes from 5 to 17~\rsun, and the density contrast increases by a factor 100. These changes show up in the increase of $n_g$ and the variation of \DPa\ from $\sim 80$~s to $\sim 60$~s. The deviations of $\Delta P$ with respect to the asymptotic behaviour are more important for less evolved models, for which the coupling between p and g cavities is  stronger.
Between RGB (C, D) and He-B (E, F, respectively) models, in addition to a significant difference in the period spacing itself, the differences between the spectra can be summarized as follows: {\it i)} in the RGB models the inertia of $\ell=1$ pressure-dominated modes (corresponding to local minima in $E$) is closer to that of the radial modes, indicating a weaker coupling between gravity and acoustic cavities compared to the He-B phase. This is also evident from the gravity-dominated modes of RGB, for which $\Delta P$ is almost constant  (as expected for pure  g-modes) except for the modes describing the minimum of inertia, while for the g-p mixed modes of the He-B model the deviation of $\Delta P$ from a constant value is more important. {\it ii)} The density of $\ell=1$ modes for the He-B model is lower than for the RGB ones.

 The amplitude of modes at different frequencies in the oscillation spectrum results from the balance between excitation and damping rates \cite{dziembowskietal2001,dupretetal2009}), nevertheless, an estimation of the relative amplitude of different modes can be provided by the normalized mode inertia \cite{houdek1999,dupretetal2009} (see also Grosjean et al. in this proceedings). 
In this framework, we can consider the modes with lower inertia as those most likely to be detected. Assuming  that the amplitude is inversely proportional to $E^{1/2}$ (see e.g. \cite{jcd2004} and references therein), and taking into account the geometrical factor for quadrupolar and dipolar modes, we build synthetic echelle diagrams for the models A-F in which the size of the point is proportional to the amplitude of the mode relative to the radial one (Fig.~\ref{fig:echelle}). The effect of evolution on the oscillation spectra is particularly strong for the ridge corresponding to $\ell=1$ modes, whose scatter decreases as the coupling between core and envelope decreases. The differences between C and E panels, corresponding to models with the same \numax\ and \LS\, led \cite{montalbanetal2010} to propose the scatter of dipole modes as a tool to distinguish between RGB and He-B objects  with global parameters typical of red clump stars. The measurement of period separation between detected $\ell=1$ modes allowed \cite{beddingetal2011} and \cite{mosseretal2011} to split up  the stars into two groups: one characterized by targets with $\Delta P > 100$~s (He-burning stars) and one with $\Delta P < 60$~s (RGB).

 In order to compare theoretical predictions and observational results, it is mandatory to define theoretical indexes as close as possible to the observational ones. Modes with lower inertia are most likely detected, but they also are the modes contributing to significant deviations from the uniform $\Delta P$  predicted by the asymptotic approximation (minima in $\Delta P$, see Fig.~\ref{fig:spec}). Taking this into account, as well as the properties of the observed oscillation spectra, \cite{montalbanetal2013} define an index $\langle \Delta P\rangle_{\rm th-obs}$, that is   a theoretical  estimation of the measurable period spacing. The predictions of $\langle \Delta P\rangle_{\rm th-obs}$  as a function of the average large frequency separation for models in the RGB and in the central He-B  phase with different chemical composition show a good agreement with the recent  observational results obtained with {\it Kepler} and CoRoT.

\section{Period spacing and core extra-mixing}

\label{sec:over}

Both the predicted \cite{montalbanetal2013} and observed \cite{beddingetal2011,mosseretal2011} period spacing show significant scatter in He-B stars. This dispersion is partly due to the different stellar masses, chemical composition and  central helium mass fraction ($Y_{\rm c}$). \cite{mosseretal2011} identified the  high-$\Delta\nu$--low-$\Delta P$ tail as corresponding to the secondary clump, but the  asymptotic and ``measurable'' values of the  period spacing contain additional information on the current and previous structure the star.

Montalb\'an {\it et al.} \cite{montalbanetal2013} have shown that  $\langle\Delta P\rangle_{\rm th-obs}$ of He-B models presents a linear relation with the mass of the He-core, with a minimum value at the transition mass (which corresponds to the minimum of He-core mass  $\sim 0.33$\msun). $M_{\rm tr}$ strongly depends on the extent of the mixed central region during the MS (see Fig.~\ref{fig:flash}, left panel) and on the chemical composition while  the values of $\langle\Delta P\rangle_{\rm th-obs}$ for RC and secondary clump stars seem only determined by $M_{\rm He}$.  Hence, the period spacing in He-B red giant stars can provide a stringent test of the central mixing during the H-MS. However, to successfully exploit this possibility, spectroscopic constraints on the chemical composition are required, together with an accurate estimate of the stellar mass: seismic constraints (other than period spacing) will be crucial in this respect.

As mentioned above, during the core He-burning phase, the high temperature sensitivity of the nuclear reactions leads to the development of a small convective core. Its size (as during core H-burning) is based on a local convection criterion, generally the Schwarzschild one. As for MS, some observational facts indicate that the extent of the central mixed region during He-B phase should be larger than that determined by local criteria. The nature of the mechanism inducing this extra-mixing as well as its extent are still a matter of debate: is it a mechanical overshooting  as proposed by \cite{bressanetal1986}, or an overshooting induced by the discontinuity of opacity generated at the external border of the convective core by the conversion of He in C and O (see e.g., \cite{castellanietal1985})  together with a semi-convective layer? The extent of the centrally mixed region has important consequences on the duration of the He-B phase, but also on the following evolutionary phases; for instance, it determines the ratio between AGB and horizontal blanch stars. Moreover, the different chemical profiles of C and O left by different kinds of mixing directly affect the oxygen abundance of later white dwarfs (see e.g. \cite{stranieroetal2003}).
 
The value of  the asymptotic period spacing depends on the central distribution of $N$, and hence on the size of the convective core. Recently \cite{jcd2011} and \cite{mosseretal2012} showed that $\langle \Delta P \rangle_a$  can be inferred from the observed period spacing of dipole mixed modes. A first comparison between \DPa\ derived for He-B models without overshooting and the values of $\langle \Delta P \rangle_{\rm a}$ derived by  \cite{mosseretal2012b} for a sample of {\it Kepler} giants,  suggests that theoretical models of low mass stars systematically underestimate $\langle \Delta P \rangle_{\rm a}$ by $\sim 20\%$ \cite{montalbanetal2013}.

Theoretical $\langle \Delta P \rangle_{\rm a}$ shows a linear relation with the radius of the convective core for models with $Y_{\rm c}$ between 0.9 and 0.1. The fit obtained from models with masses between 0.7 and 4.0~\msun, and metallicity $Z=0.02$ is : $\langle \Delta P \rangle_{\rm a}=17.35+1.06176\times 10^4\times R_{\rm cc}(R_\odot)$ \cite{montalbanetal2013}.

Mixing processes during the MS phase do not affect the central regions of low mass stars in He-B phase since the mas and the physical properties of the helium core are fixed by the onset of He-burning in degenerate conditions. To increase $\langle \Delta P \rangle_{\rm a}$ we should  decrease the contribution of $N$ in the innermost regions of the star, which can be easily done by extending the adiabatically stratified and chemically homogeneous core. So, models computed with central extra-mixing during the He-B evolution show an increase of the period spacing as the size of homogeneous core increases, and the values of \DPa\ follow the same linear relation. The exact $\langle \Delta P \rangle_{\rm a}$  value  depends on the extent of the extra-mixed region, but also on the details of the mechanism at the origin of different $N$ profiles: a mechanical overshooting with adiabatic temperature gradient in the extra-mixed region, or an induced overshooting with a semi-convective layer outside the convective core. The latter will provide, for the same extent of the adiabatic region, a smaller period spacing than the former. 

{\it To sum up, dipole mode period spacings in RGs allows us to constrain the extent of extra-mixing during core H- and He-burning phases. The quasi-linear relation between the ``observed''  period spacing in secondary clump stars and their He-core mass contains information about the extent of the mixed region during the H-MS. Currently, seismology allows us to derive mass and secondary clump membership for a huge number of field RGs. The comparison between the theoretically expected $\langle\Delta P\rangle_{\rm th-obs}$ distribution for synthetic stellar populations with observed distributions will shed some light on the mixing processes during the H-MS. Moreover, $\langle\Delta P\rangle$ for RC stars will provide promising indices about the extent of the mixed central region during core He-burning in low-mass stars.}

\section{Stellar rotation from dipole modes in red giants}
\label{sec:rotation}

Stellar rotation lifts the frequency degeneracy of non radial modes with the same degree ($\ell$) and different azimuthal order ($m$) producing multiplets of modes with $(2\ell+1)$ components. For rotation velocities low enough with respect to the oscillation frequency the first order approximation is still valid and the frequencies of the multiplet are symmetrically distributed around the non-rotating value ($\omega_0$) (see Ouazzani contribution, these proceedings, for a detailed discussion): 

\begin{equation}
\omega_{n\ell m} = \omega_0 \pm m\,\int_0^R K_{n \ell}(r)\,\Omega(r)\,dr, 
\label{eq:rot}
\end{equation}

\noindent where $\Omega$ is the angular velocity, and $K_{n\ell}$  the rotation kernel, which depends only on the radial and horizontal components of the eigenfunction ($\xi_r$ and $\xi_h$ respectively) and on the degree $\ell$:

\begin{equation}
K_{n\ell}(r)=\frac{\left(\xi_r^2+(\ell+1)\ell \xi_h^2-\xi_r\xi_h-\xi_h^2\right) r^2\rho}{\int_0^R(\xi_r^2+(\ell+1)\ell\xi_h^2)r^2\rho\,dr}.
\end{equation}

In the case of uniform rotation ($\Omega(r)=\Omega_s$), the rotational splitting between components of the same multiplet can we written as: $\Delta \omega _{\rm rot} = (1-C_{n\ell}) \Omega _s$,  where $C_{n \ell}$ is the so-called Ledoux constant. $C_{n\ell} \simeq 1$ for pure acoustic modes,  $C_{n\ell} \simeq 0.5$ for dipole pure g-modes, and is comprised between these two limits for dipole g-p mixed modes. Therefore, in the conditions of validity of the first order approximation and of uniform rotation,  the rotational splitting of gravity-dominated  modes should be smaller than that of pressure-dominated modes.  However, \cite{becketal2012} derived, for a red giant of 5~\rsun, a rotation splitting that increases as the g-character of the mode increases, producing a "V" shape variation of rotation splitting around the pressure-dominated mode. This behavior has been interpreted as a proof of a differential rotation inside red giant stars, with the core rotating 10 times faster than the surface.
Rotation splitting of mixed modes has also been measured in other post-MS stars: sub-giants \cite{deheuvelsetal2012}, another red giant of 5~\rsun\ stars (Di Mauro {\it et al.} this proceedings) and red giants in different evolutionary states (\cite{mosseretal2012} and these proceedings). The explanation of these new results is a challenge for the current models of stellar evolution including rotation (\cite{eggenbergeretal2012}).

\begin{figure}
\resizebox{\hsize}{!}{{\includegraphics{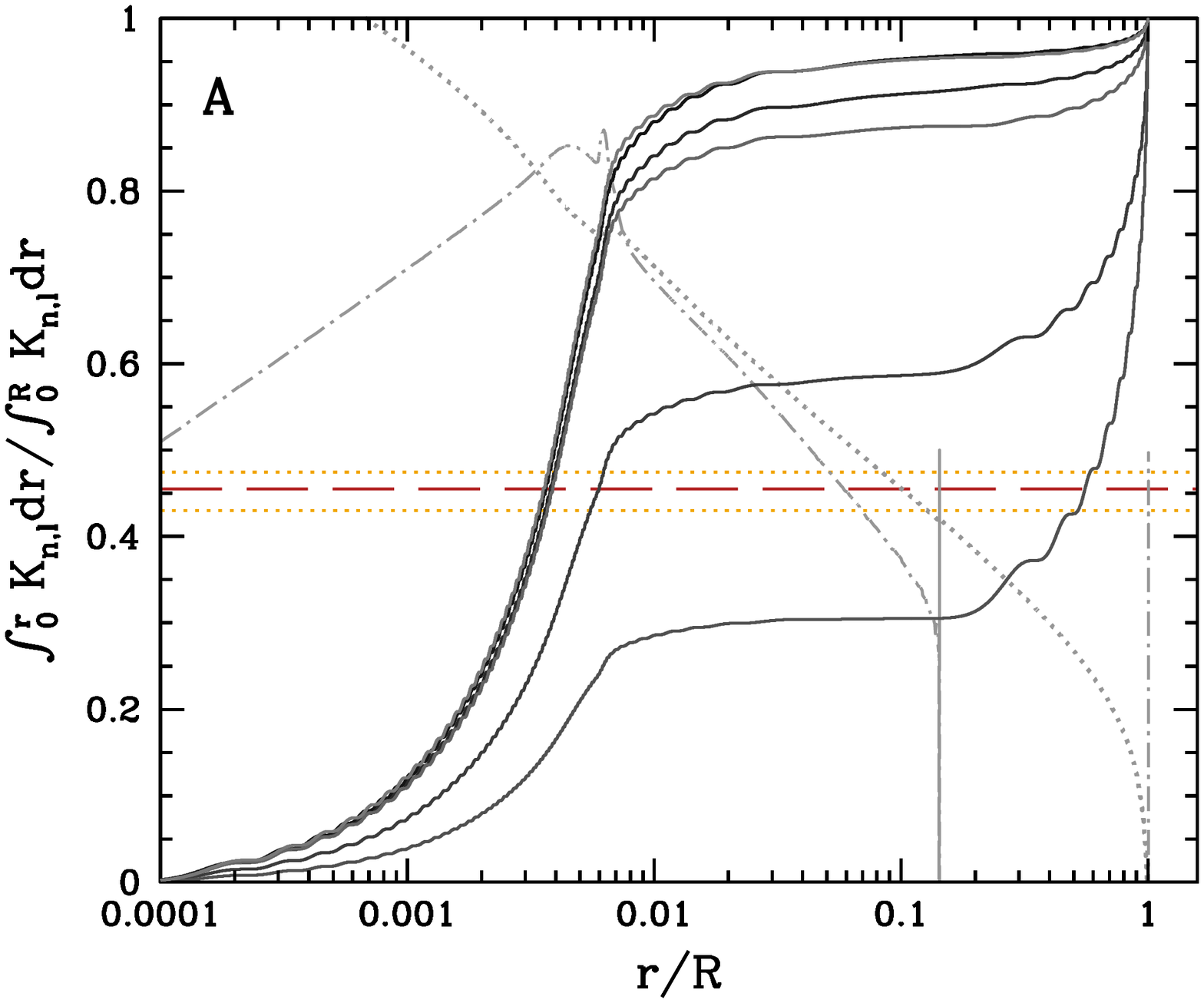}}{\includegraphics{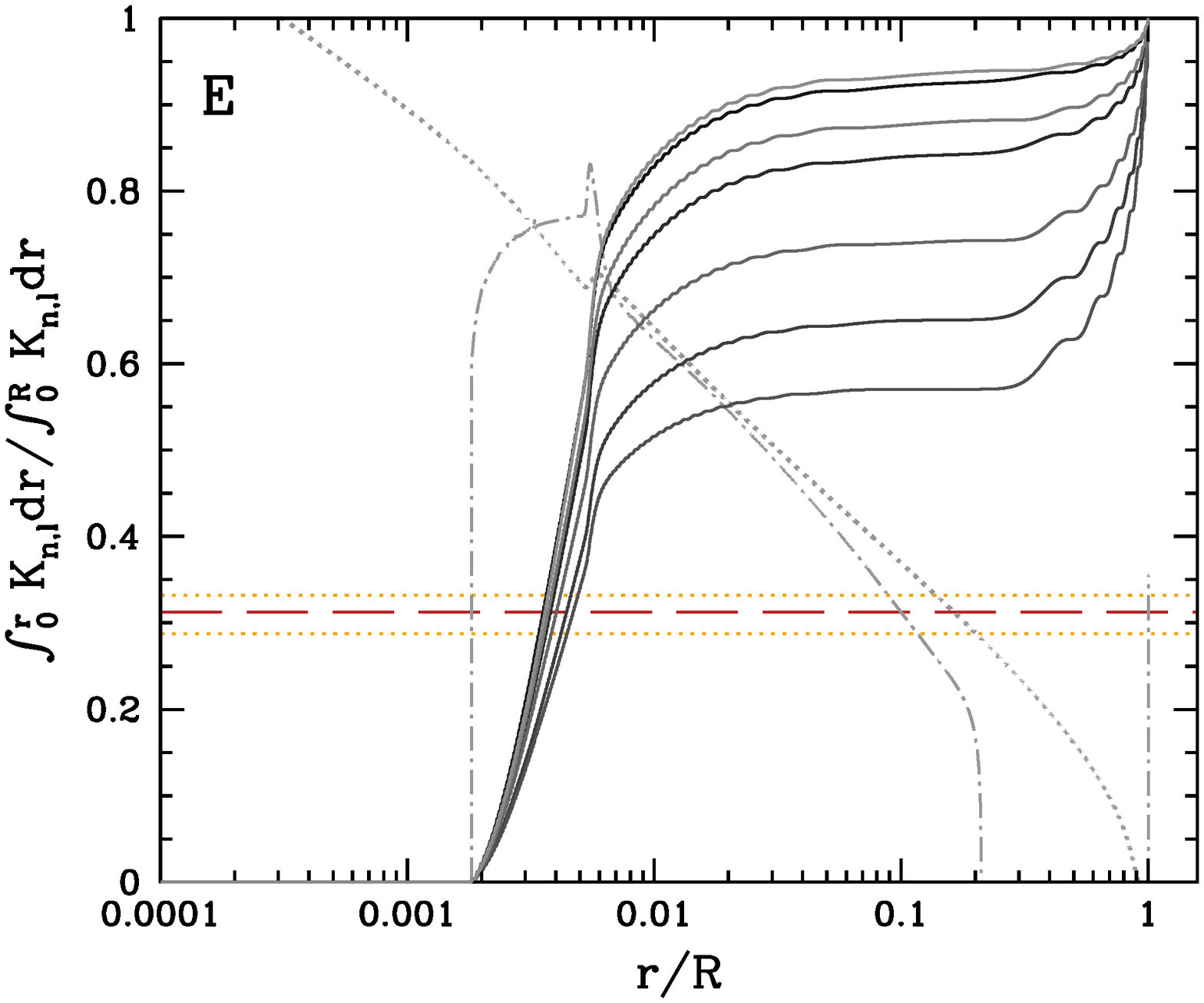}}}
\caption{ Partial integrals of normalized rotation kernels  as a function of the relative radius for dipole modes belonging to the same "dipole forest". $N$ (dash-dotted light gray curves) and $S_\ell$ (dotted light gray curves). Left panel correspond to the 1.5~\msun\ RGB model labeled A in Figs.~\ref{fig:hr}, ~\ref{fig:spec}, ~\ref{fig:echelle}. The modes represented are those between 183.6 and 195.25 $\mu$Hz (see Fig. ~\ref{fig:spec}, panel A). Right panel corresponds to the RC model labeled E, and the rotation kernels corresponds to dipole modes between 37.24 and 39.04~$\mu$Hz.}
\label{fig:rot}
\end{figure}

Since the kernel follows the properties of the eigenfunction, it reflects also the sensitivity of the mode to the angular velocity of a given region of the star. In Fig.~\ref{fig:rot}  we plot the partial integrals of the normalized rotation kernels illustrating the contribution from different regions of the star to the rotational splitting of likely observable dipole modes. 
We also show the profiles of $N$ and $S_\ell$ to clearly relate the behavior of the kernels with the stellar structure features.  It is worth noticing that for a given ratio between surface and central angular velocities, the difference of rotation splitting undergone by  pressure-dominated modes  and that of mixed modes belonging to the same "dipole forest"  is smaller for red-clump models than for models on the RGB. In the RC model, the central regions contributes to the splitting by 55\% for the pressure-dominated modes, and by 90\% for the gravity-dominated ones, while for the RGB model, this contribution varies from 30 to 90\%.

%\bibliographystyle{epj}
%\bibliography{dP}

\end{document}